# High entropy alloy superconductors - status, opportunities and challenges

Liling Sun[1,3] and R. J. Cava[2]

[1]*Institute of Physics and Beijing National Laboratory for Condensed Matter Physics, Chinese Academy of Sciences, Beijing 100190, China*

[2]*Department of Chemistry, Princeton University, Princeton, New Jersey 08544, USA*

[3]*University of the Chinese Academy of Sciences, Beijing 100190, China*

High entropy alloys (HEAs) are a recently-opened research area in materials science and condensed matter physics. Although 3*d*-metal-based HEAs have already been the subject of many investigations, studies of HEA superconductors, which tend to be based on 4*d* metals, are relatively fewer. Here we provide a short update of the progress made in studies of superconducting HEAs. We aim to summarize their current status and describe some of the key factors that appear to influence their superconducting transition temperatures and properties, including crystal structure, atomic makeup, valence electron count, molar volume and mixing entropy. Many opportunities and challenges remain for expanding our knowledge of HEA superconductors, finding new types of HEA superconductors, and their potential for applications; these are also briefly discussed.

High entropy alloys (HEAs) are relatively new class of materials that have attracted significant research interest in materials science and engineering. Simply speaking, these are alloys or compounds that consist of several different types of randomly mixed constituent atoms and thus display a high degree of disorder, i.e. a high configurational entropy. Although an exact classification of what constitutes a "high" versus a "medium" versus a "low" entropy system depends on which of the historical definitions is used, a single general principle is maintained in all cases, and that is that an HEA phase displays extreme atomic disorder due to the mixing of several different kinds of atoms in significant fractional quantities on its crystal lattice. Two different schemes for classifying such materials have come into being historically. By one classification, HEAs are alloys that simply have between 5-35% of each of 5 or more elements [1-5]. By another classification scheme, the calculated idealized entropy of mixing is the determining factor. This second classification scheme works because the ideal entropy of mixing is easily calculated as $\Delta S = -\Sigma x_i \ln x_i$. (Note that the thermodynamic stability of a material actually depends on ΔG, its free energy of formation, and so its thermodynamic stability criterion includes, among other factors, the terms $\Delta H_{mixing}$ and $T\Delta S_{mixing}$, $\Delta H_{mixing}$ is not easily determined, and so, for functionality purposes, the ideal entropy of mixing is used.) To avoid the difficulty of having to know the ΔH values, whether the system is actually an ideal solution, and anything about the distribution of different atoms on different kinds of sites in crystal structures where more than one type of crystallographically independent site is possible for the atoms, (e.g. for the "CsCl" and "α-Mn" structure types encountered for HEA superconductors) a simple criterion based on the calculated ΔS of mixing for an ideal solution is used. The distinctions between “high” “medium”, and “low” entropy alloys based on this entropy of mixing criterion are practical, in the end, and harken back to the other historical definition. Generally speaking, low entropy alloys have idealized entropies of mixing of (magnitude) less than -0.69R (R is the ideal gas constant), the idealized entropy of mixing of a simple equimolar binary alloy, medium entropy alloys have idealized entropies of mixing between -0.69R and -1.59R, and high entropy alloys have idealized entropies of mixing with magnitudes of -1.60R or larger (the idealized entropy of mixing for an equimolar 5 atom solid solution). The different historical definitions can sometimes lead to differences in what an alloy is called, but not in the general principle - a great deal of disorder and configurational entropy is present. Even an equimolar quaternary alloy will not satisfy either of the above criteria for HEA status (the idealized entropy of mixing is -1.39R) but it is clearly extremely disordered; it can be classified as a MEA.

One of the most striking features of HEAs is that the atoms randomly arrange themselves on the crystallographic positions of simple lattices (Fig.1a), thus resulting in alloys possessing high configurational entropy [2-7]. Fig.1b shows a schematic of the general compositional difference between HEAs and the usual functional alloys. The closer the alloy composition to the central region, the larger the value of the entropy of mixing per mole (The thermodynamics of superconducting HEAs is the same as that of non-superconducting HEAs, which has been treated extensively [7]). The crystallographic and chemical character, and extreme atomic disorder present in HEAs suggests to us that these materials can be considered as glasses on a crystalline lattice. Until now, more than ninety different types of as-cast HEAs have been found [2], which mainly form in three types of simple crystal structures: body-centered cubic (BCC), hexagonal closest-packed (HCP) and face-centered cubic (FCC) [2,6,8]. HEA (or medium entropy) superconductors currently known are found in BCC, α-Mn, CsCl and HCP crystal structures.

In line with their high randomness on simple lattices, HEAs display mechanical and physical properties that have motivated their continuing study. For example, they can possess high fracture toughness at cryogenic temperatures – their low stacking fault energy promotes twinning in HEAs [9, 10], and the sluggish diffusion of their constituent elements allows HEAs to display high strengths (>1 GPa) even at temperatures as high as 1200° C [11]; further, they can display high specific strengths – the unique crystal structure of the HEAs is known to provide compressive yield strengths up to 1.02 GPa at ambient temperature [12, 13]; and, finally, they are reported to be more resistant to general corrosion than is 304 stainless steel in a de-aerated 1 N $H_2SO_4$ solution, for example [14]. These types of properties may be of interest for HEA superconductors as well.

Most HEAs studied have involved primarily constituents from the 3*d* series of metals, but in 2014, the first HEA superconductor, consisting primarily of 4*d* and 5*d* series elements, was found [15], revealing a new facet of the capabilities of these materials. A variety of studies have been performed on HEA superconductors since that time [7, 16-26], in part because the combined electronic and mechanical properties found in HEAs suggest their potential for practical applications. Being in the relatively early days in the study of HEA superconductors, it is possible to consider all of the known ones and present a reasonably complete snapshot of HEA superconductors at this point in time.

In this update, we focus on the variations of HEA superconductors presently known, and the progress in their study, especially highlighting some fundamental issues related to their superconducting transition temperatures ($T_C$). Also, we explore the opportunities for finding new types of HEA superconductors and their potential applications as well as current challenges to our understanding of their superconducting behavior.

## 1. The types of HEA superconductors and research progress

To fabricate HEA superconducting alloys, stoichiometric amounts of the metals are typically arc melted using high currents (at T >2,500° C) in an argon atmosphere. Usually, the samples need to be melted several times and turned over each time to ensure optimal mixing of the constituents. The fast cooling rates generally required to form these alloys are consistent with their entropic stabilization. In these ways their fabrication is similar to what is known for 3*d*-based HEAs, with more diverse synthesis methods reported in those cases [7,16]. Up to the present time, thin films of superconducting HEAs have not been reported. Powder X-ray diffraction is the typical first-order characterization method for determining the phase assemblage and average structures of HEAs, superconducting and otherwise. This method is useful for determining the crystal structure of an HEA averaged over all local configurations of atoms, but for finer scale structural and microstructural characterization, high-resolution transmission electron microscopy (HRTEM) and its accompanying elemental analysis is very useful method for detecting the fine-scale compositional distribution and microstructure in these materials. Applying this method, the HEA superconductor $(TaNb)_{0.67}(ZrHfTi)_{0.33}$ has been studied [7]. The results clearly indicate that the BCC structure of this HEA is homogeneous even at the nanoscale. The actual compositions that form single phase HEAs are limited – any general mixture of elements will not do – multiple phase products are the rule rather than the exception. This is an important factor for researchers to consider when determining the characteristics of HEA superconductors.

Among the HEAs, four types of superconductors have been discovered so far, as listed in Table 1, although some of them may be classified as medium entropy alloys. The type-A HEA superconductors consist of only metals on the left side of the transition metal region of the periodic table. Noteworthy in this group are the Ta-Nb-Hf-Zr-Ti and Nb-Zr-Ti-Re superconductors [21-24, 27, 28]. This type of HEA superconductor crystallizes on a small unit cell BCC lattice ($a_0$ ~ 3.2 angstroms) in space group *Im3m*, and exhibits superconducting transition temperatures ($T_C$) between 4.0 and 9.2 K. Their highest upper critical field, $\mu_0H_{c2}(0)$, is about 11.7 T. Figure 2 shows an example of a superconducting transition in one of these HEAs, demonstrated by the temperature dependence of resistance, heat capacity, and magnetic susceptibility measurements. These are so far the most frequently studied HEA superconductors. Type-B HEA superconductors mainly consist of the early transition metals, but crystallize on a relatively complex lattice [23]. The HEAs with high *x* values crystallize on a (larger BCC) α-Mn type lattice, while the HEAs with lower *x* values crystallize on a mixture of smaller BCC and HCP lattices. Examples of this type of HEA superconductor include $(ZrNb)_{0.3}(MoReRu)_{0.7}$, $(HfTaWIr)_{0.7}Re_{0.3}$ and $(HfTaWPt)_{0.5}Re_{0.5}$ (the number outside of the round brackets represents the sum of the chemical compositions of the elements in parentheses, for instance, $(ZrNb)_{0.3}$ refers to $Zr_{.15}Nb_{.15}$). The latter two HEAs are made from only 5*d* elements, and the last material has an ideal $\Delta S_{mixing}$ = 1.39R (i.e. lower than 1.6R) and one of the constituents (Re) is present at higher than the 35% fraction used to define HEA limits in the alternative classification. The $T_C$'s of type-B HEA superconductors appear to be lower, from 1.9 to 5.7 K, and the highest known $\mu_0H_{c2}(0)$ is about 7.9 T. Type-C HEA superconductors are composed of transition metals on the left side of the *d*-block combined with transition metals on the right side of the *d*-block. Examples are Sc−Zr−Nb−Ta−Rh−Pd superconductors [16]; these materials have a small unit cell ($a_0$ ~ 3.2 angstroms) with a CsCl-type structure in space group *Pm3m*. The transition temperature $T_C$ of type-C HEA superconductors varies from 3.9 to 9.3 K, and the largest $\mu_0H_{c2}(0)$ is about 10.7 T. We note that for type-B HEA superconductors, which have the α-Mn type lattice, although the crystallographic unit cell is larger than those for the type-A and type-C HEA superconductors, the crystal structure is based on the packing of small metal clusters and therefore the elementary atomic building blocks are not significantly larger than for the "simpler" type-A and type-C cases [16, 23]. Type -D HEA superconductors were found recently [29], this type of HEA superconductor is $Re_{0.56}Nb_{0.11}Ti_{0.11}Zr_{0.11}Hf_{0.11}$, which crystallizes on a HCP lattice. Its superconducting transition temperature and lower critical field are 4.4 K and 2.3 mT, respectively. For this material, the ideal $\Delta S_{mixing}$ = 1.29R (i.e. lower than 1.6R) and one of the constituents (Re) is present at higher than the 35% fraction used to define HEA limits in the alternative classification.

Since the first example of an HEA superconductor was discovered, a variety of experimental studies have been made to look for new HEA superconductors and to understand the underlying physics of this family of materials. Detailed theoretical studies of the electronic structures are existent [19, 21] – they are challenging due to the extreme atomic disorder present. Even if conventional, however, the superconductors display some remarkable properties asking for explanation via standard electron-phonon coupling models for superconductivity and also suggesting their potential for future applications, especially under extreme pressure conditions. Thus, we argue, if a broader view than the usual "how high is Tc" view is applied, these materials are worthy of continuing study.

*1.1 A correlation between valence electron count (VEC) VEC and $T_C$*

A critical parameter that can affect the stability of a crystalline solid solution phase in the absence of an overriding atomic size effect is the valence electron count (VEC) [30], which for the HEAs is best reported as the number of valence electrons per atom. Experimental results have shown that the VEC of the HEA superconductors plays a crucial role in determining their superconducting $T_C$ as well as the stability of the material. $T_C$ for the HEA superconductors strongly depends on the VEC in a fashion reminiscent of the Matthias plot for $T_C$ vs. electron count in simple binary transition metal alloys [7, 16, 17]. The superconductivity of type-A HEA superconductors $(TaNb]_{1-x}(ZrHfTi)_x$ emerges in the VEC range of 4.3-4.8. $T_C$ increases with VEC and reaches a maximum (7.3 K) near the value of 4.7 (light orange region in Fig. 3). Critically, so far the stability of materials with the small-unit-cell BCC structure in the type-A superconductor family, for materials prepared by fast cooling from high temperatures (there are no current reports on the effects of very high cooling rates), is limited at both the high VEC side and the low VEC side. Although it is possible to clearly see the trends in superconducting $T_C$ with electron count, a wider range of stable VEC materials of type-A, perhaps accessible through very fast cooling rates, would be a welcome addition to the knowledge base of this superconducting material family.

Variations in $T_C$ that are induced by atomic substitutions, via valence VEC changes in particular, seem to be a universal behavior in HEA superconductors. A VEC dependence of the superconducting transition temperature is also observed in type-C HEA superconductors such as $(ScZrNb)_{1-x}(RhPd)_x$ [16]. The superconductivity in this case appears in the VEC range of 5.9-6.3 and the maximum $T_C$ (9.3 K) presents at the VEC ~5.9.

When considering the $T_C$'s of all three types of HEA superconductors a full electron count trend can be observed. This is shown in figure 3. The small unit cell BCC type-A and type-C CsCl-type HEA superconductors do seem to generally follow the two-peak character seen for the simpler binary alloys near VECs of 4.7 and 5.9, although not in detail, while the more complex α-Mn type-B HEA superconducting alloys do not seem to follow the general trend seen for the other two systems. Although the small unit cell BCC and CsCl crystal structures are highly related, (the latter has different distributions of atoms in the ½ ½ ½ and 000 positions while the former has the same distributions of atoms in the same positions) their atomic makeups are quite different and, most critically, their stabilities are over limited ranges of VEC, thus frustrating a straightforward comparison. It is a clear opportunity for future research to find an HEA superconducting system that is stable over a very wide range of VEC and study its superconducting properties.

Comparison of the superconducting transition temperatures of the crystalline transition metals and their alloys [31], amorphous vapor deposited films [32], and the type-A BCC HEA superconductors reveals that the $T_C$ of the HEA superconductors is clearly lower than that of the crystalline alloys and follows a monotonically increasing trend of values higher than those of the amorphous alloys. The superconducting transition temperature of the HEA lies between the binary crystalline alloy and amorphous material at the same electron count. In view of these results, it is evident that when comparing crystalline binary alloys, HEAs, and amorphous materials at the same electron count, the degree of crystallinity has a significant effect on $T_C$. Although a detailed physics comparison among chemically and electronically similar representatives of amorphous, HEA and binary superconductors has not yet been made, this suggests to us that the smearing of the electron energy (*E*) versus electron wavevector (*k*) relations and thus the broadening of the electronic density of states due to disorder is the likely cause for this behavior. The results reveal that more perfect crystallinity has a positive impact on $T_C$ in this alloy group.

### 1.2 *Influence of atomic substitutions with the same VEC on $T_C$*

Experimental results have shown that elemental replacements at the same VEC influence the superconducting $T_C$ of HEAs significantly. The first example is illustrated by studies of isoelectronic replacements using Mo+Y, Mo+Sc, and Cr+Sc mixtures for the valence electron count 4 and 5 elements in the type-A BCC Ta-Nb-Zr-Hf-Ti HEA superconductor [17] (Mo and Cr have VEC 6, Ta, Nb, and V have VEC 5, Hf, Zr and Ti have VEC 4 and Y and Sc have VEC 3). The results show that the superconducting $T_C$ strongly depends on the elemental makeup of the alloy, as shown in Table 2. The replacement of Nb or Ta by an isoelectronic mixture lowers the transition temperature by more than 60%, while the isoelectronic replacement of Hf, Zr, or Ti has a limited impact on $T_C$. Efforts to alloy aluminum (Al) into the nearly optimal electron count $(TaNb)_{0.67}(ZrHfTi)_{0.33}$ type-A HEA superconductor have also been made [17]. The electron count dependence of the maximum superconducting $T_C$ for the $[(TaNb)_{0.67}(ZrHfTi)_{0.33}]Al_x$ series is found to be 4.7, same as that of the $(TaNb)_{1-x}(ZrHfTi)_x$ HEA solid solution. For an aluminum content of $x = 0.4$ the high-entropy stabilization of the simple BCC lattice breaks down, a multiple phase material is observed, and superconductivity is not seen above 1.8 K.

The composition dependence of the superconducting transition temperature in the $(ScZrNb)_{1-x}(RhPd)_x$, type-C HEA superconductors with the CsCl structure, has also been investigated [16], although primarily to observe the effect of VEC on $T_C$. Studies that directly address the effect of elemental make-up on $T_C$ at the same VEC in the type-C HEA superconductors would be of future interest.

Elemental Nb appears to be an important ingredient for determining the $T_C$ of the HEA superconductors; it appears that $T_C$ is optimized by the presence of elemental Nb even when the same VEC is accessible through the use of other elements. In the type-A alloys $Nb_{67}(HfZrTi)_{33}$, $(NbTa)_{67}(HfZrTi)_{33}$ and $(TaNbV)_{67}(HfZrTi)_{33}$, for example, $T_C$ is 9.2 K in $Nb_{67}(HfZrTi)_{33}$ with the highest Nb content (close to the $T_C$ value (9.3 K) of pure Nb), 7.6 K in $(NbTa)_{67}(HfZrTi)_{33}$ with an intermediate Nb content and 4.3 K in $(TaNbV)_{67}(HfZrTi)_{33}$ with a lower Nb content [7]. When Nb is fully absent for HEA materials with the same VEC, such as for $[Ta(Mo_{0.5}Sc_{0.5})V]_{67}(HfZrTi)_{33}$, no superconductivity is observed above 1.9 K. Thus for HEA alloys at the same VEC with (essentially) the same BCC unit cell size, $T_C$ is substantially improved when Nb is present. Because the density of state (DOS) values are expected to be virtually the same at the same electron count and cell volume in these highly disordered systems, we interpret this observation to mean that Nb is an important constituent in intermetallic superconductors even when present at a relatively low level because it somehow enhances the electron-phonon coupling.

### *1.3 Influence of mixing entropy on $T_C$*

A unique characteristic of the HEAs is the high mixing entropy due to the random arrangement of the elements on the lattice [2, 7, 15]. Consequently, a crucial question emerges: what is the relation between the mixing entropy and the superconductivity? In other words is the mixing entropy an important factor to determine the $T_C$ of the HEA superconductors? There have been almost no studies to address this question explicitly. One study [7] looked at the correlation between the mixing entropy and $T_C$ for the Ta-Nb-Hf-Zr-Ti type-A HEA superconductors and found that increasing the mixing entropy does not have a conclusive effect on the $T_C$, as shown in Fig.4. These results suggest that the central role of the high configurational entropy in HEA superconductors is to stabilize their high symmetry crystal structures instead of determining their $T_C$s. There is also partial information on the effect of increasing entropy on the upper critical fields, which, in a general way appear to increase with increasing entropy in type-A HEA superconductors. Future work would be required to establish this apparent trend more solidly.

*1.4 Influence of volume shrinkage on the $T_C$*

It is well known that compression of superconducting materials can lead to changes of both the crystal and the electronic structure, the latter often due to the shortening of the interatomic distances present. Thus, high pressure has been widely adopted as a control parameter to tune the superconductivity of materials. This kind of experiment is considered to be cleaner, for example, for determining the effect of lattice size on $T_C$ than the case of "chemical pressure" where a smaller atom is substituted for a larger one, because effects that are due to changing the mixture of atoms present are not introduced. The examples include pressure-induced enhancement of the transition temperature $T_C$ in copper-oxide and iron-pnictide superconductors [33-38], reemergence of superconductivity in the alkaline iron selenide [39] and heavy fermion superconductors [40], pressure-induced superconductivity in $H_3S$ and $LaH_{10}$ [41-41] and in normally non-superconducting elements [45,46]. Therefore, to know what happens for pressurized HEA superconductors is of great interest. Using diamond anvil cells, the superconducting behavior of the type-A $(TaNb)_{0.67}(HfZrTi)_{0.33}$ HEA at high pressures was investigated [18]. Surprisingly, the zero resistance superconductivity of this material exists continuously from 1 atm to a pressure as high as 190.6 GPa (about 2 megabars), a pressure like that within the outer core of the earth. $T_C$ increases with pressure, exhibiting a slow increase from its ambient-pressure value of ~7.7 to 10 K at ~60 GPa. On further increasing the pressure, $T_C$ remains almost unchanged up to ~190 GPa. The extraordinarily robust zero-resistance superconductivity in this HEA superconductor is extremely unusual and virtually unique among known superconductors (later, the same group observed similar behavior in a multiple phase NbTi superconducting filament extracted from a commercial alloy cable [47].) The upper critical fields ($Hc_2$) at zero temperature for the pressurized $(TaNb)_{0.67}(HfZrTi)_{0.33}$ type-A HEA superconductor are estimated as $Hc_2(0) = 4$ T at 100 GPa and 2 T at 179.2 GPa, although the latter is a relatively low upper critical field for an ambient pressure material, it has to be considered as remarkable for a material at nearly 2 million atmospheres of pressure – these results thus indicating that HEA superconductors can display properties that may make them applicable under high pressure conditions.

A similar phenomenon was also found in the type-C HEA superconductor $(ScZrNbTa)_{0.6}(RhPd)_{0.3}$ under pressure [48]. $T_C$ is found to increase upon compression but saturates starting at pressure ~ 30 GPa. With increasing pressure from 30 GPa to 90 GPa, the transition temperature $T_C$ stays almost constant. These results suggest that HEA superconductors have a universal pressure-$T_C$ phase diagram, as shown in Fig.5.

High-pressure synchrotron X-ray diffraction (XRD) measurements on the type-A and type-C HEA superconductors have also been performed [18,48]. The experimental results show that the superconducting HEAs do not undergo structural phase transitions up to pressures of ∼100 GPa, but that their volumes are compressed by ∼30%. Thus the BCC and CsCl structures are highly stable under compression. Why $T_C$ stays nearly constant after the molar volume is reduced by ~30% is an interesting issue and remains an open question worthy of future study. Moreover, it would also be of interest to know how high a pressure would be needed to cause the superconducting HEAs to become structurally unstable. We assume that this will happen when the crystal structure is destroyed or collapsed by external pressure, perhaps, for the BCC and CsCl-type materials, through a transformation to an FCC of HCP lattice.

Like their 3*d*-based cousins, the superconducting HEA materials appear to have unusual mechanical properties [1-6]. Although an important branch of study in the materials science of 3*d* HEA materials, there are as yet no formal studies published for HEA superconductors but anecdotally at least the usual methods used for grinding samples to obtain powders for studying diffraction often do not work for these materials because the fracture toughness of the superconducting HEAs is too high. More extreme measures such as rolling foils or snipping small pieces from arc-melted buttons must be used to prepare samples for this kind of analysis. The fact that type A and type C HEA superconductors have survived without fracturing up to around 200 GPa in high pressure experiments is another indication of their high fracture toughness.

## 2. Opportunities

Some of the opportunities for future study of the known HEA superconductors have been described in the previous sections. More generally, though, possessing the distinguished properties of excellent specific strength and superior mechanical performance at high temperatures (expected based on the well-studied 3*d* analogs), together with extraordinary robustness of superconductivity at extreme conditions, HEA superconductors are expected to offer many opportunities for future scientific inquiry.

Recently, for example, a new superconductor has been presented that incorporates a large amount of chemical disorder in a layered superconductor [25]. Layered superconductors have a crystal structure based on alternate stacks of electrically conducting layers, such as $CuO_2$, FeAs, and $BiS_2$ layers, with blocking layers, which are often composed of metal oxides. The innovation of the above work is that the highly disordered intermediate layer used can be considered an HEA. The performance of the resulting layered superconductor shows that it has better superconductivity than is usually observed in this family [46,49,50]. The mechanical properties were not reported, and it is not known whether the introduction of an HEA-like structure into one part of an otherwise brittle material will impact its mechanical properties significantly. The introduction of an HEA into the blocking layers in layered superconductors provides a path to explore new superconductors with the possible combined advantages of high superconducting transition temperature and good mechanical properties.

In addition, some of the novel phenomena found in the HEA superconductors, such as the continuous existence and strengthening of superconductivity against large volume shrinkages, provide new opportunities for a better understanding of electron-phonon coupled superconductivity, although it may be that that current models for superconductivity may not be up to the challenge of explain this observation. These systems may thus help to explore the effect of molar volume on superconductivity more generally.

Being "new materials" people, we are particularly interested in the opportunities that HEAs provide for the discovery of new superconductors. It seems to us that all the possible simple alloy structure types have not yet been explored from the HEA perspective, and also that nothing intrinsically seems to limit their $T_C$ to below the 10 K range. The next few years, we expect, may bring some interesting new HEA superconducting materials types to the fore.

Finally, although we are not theorists, it seems to us that HEAs may provide opportunities for research for theorists who are interested in conventional mechanisms for superconductivity. An example of such an opportunity may lie in calculating in a rigorous way how the electron *E* versus *k* relations and resulting Density of Electronic States (DOS) is impacted by increased disorder at the same valence electron count, *i.e.* for the BCC HEA superconductors with the same VEC how do the *E* vs. *k* relations and the DOS change on going from a binary to a ternary to a quaternary to a pentinary alloy? Also, why does the amount of Nb present matter so much in pentinary alloys at the same VEC and cell volume? Is it really, as we guess, that the electron-phonon coupling increases as a function of Nb content? If so, then why?

## 3. Summary and perspective

The superconducting behavior of HEAs is distinct from copper oxide superconductors, Fe-based superconductors, conventional alloy superconductors and amorphous superconductors, suggesting that they can be considered as a new class of superconducting material. Understanding the microscopic physics of superconductivity in the HEA superconductors is the subject of active ongoing studies, both theoretical and experimental, although not widely. What is known about superconducting HEA materials, as outlined in this short update, can be summarized below:

(1) Until now, four types of HEA superconductors have been discovered. The type-A HEA superconductors (for example, the Ta-Nb-Hf-Zr-Ti superconductors) consist of the early transition metals and crystalize on a small unit cell BCC lattice. Type-B HEA-superconductors (for example, the $(HfTaWIr)_{1-x}Re_x$ superconductors, x<0.6) mainly consist of the 5*d* transition metals, and crystallize on a larger-unit-cell cluster-based BCC lattice. Type-C HEA superconductors (for example, the Sc−Zr−Nb−Ta−Rh−Pd superconductors) are composed of the early transition metals and the late transition metals and crystallize on a small cell CsCl-type lattice. Type-D HEA superconductors (for example, the $Re_{0.56}Nb_{0.11}Ti_{0.11}Zr_{0.11}Hf_{0.11}$ superconductor) crystallize on a HCP lattice.

(2) The HEA superconductors that crystallize on the small cell BCC or CsCl-type lattices have the highest transition temperatures. All $T_C$s so far are limited to the sub-10-K range.

(3) The superconducting $T_C$s of the HEAs so far found are intermediate between those of amorphous alloys and simple binary alloys, at a fixed VEC following a trend of increasing $T_C$ with decreasing disorder.

(4) Although the type-A and type-B HEA superconductors have highly disordered atoms on simple lattices, the effects of elemental makeup and valence electron count on their physical properties are significant. For the latter property, the $T_C$ values mimic the classic

Mathias behavior seen for binary alloys, although not in detail, and are limited by the chemical stability.

(5) Increasing the configurational entropy by adding elements has no decisive effect on the $T_C$ of the HEA superconductors, but can stabilize their cocktail-like crystal structures.

(6) Under pressure, the HEA superconductors exhibit robust superconductivity against volume shrinkage without structural phase transitions, with the common feature that $T_C$ saturates at a constant optimal value at a critical pressure that changes from system to system.

(7) Although to our knowledge there are no detailed studies yet published, the evidence presented in publications centered on the electronic properties of these materials indicates that HEA superconductors, like their other HEA counterparts, have a high fracture toughness.

(8) The HEA superconductors currently known display type-II behavior. Though the upper critical magnetic fields of the current HEA superconductors are not as high as those of NbTi or $Nb_3Sn$, which are employed in fabrication of the majority of commercial superconducting magnets at present, we expect that future superconducting HEAs may be good candidate materials for the fabrication of the superconducting magnets.

Some further questions and issues that may be suitable for future work are proposed as follows:

(1) Why do the VEC and chemical substitutions affect the $T_C$ of HEA superconductors? Understanding the interplay among electron count, lattice structure, and the substitution of different elements at the same electron count on $T_C$ may help to find new HEA superconductors with higher $T_C$.

(2) What are the differences and similarities in the superconducting properties of HEAs and binary transition metal alloys? Can HEA superconductors be understood as primarily extremely disordered versions of superconducting binary alloys or is there a more fundamental difference?

(3) At a critical pressure that is different for different systems, the $T_C$ of the HEA superconductors stays almost constant while their volumes change dramatically. Why is that the case? Does the $T_C$ of the HEA superconductors depend on the phonon and band structure of the materials? Can that be measured experimentally?

(4) Theoretical study of the electronic structures and phonon spectra of HEA superconductors are relatively few. What do the electronic structures and lattice vibrations of HEAs actually look like? Do they differ from those of amorphous metal superconductors or quasicrystal superconductors? What kinds of theoretical treatments of highly disordered materials will be of interest when applied to HEA superconductors?

(5) "What is the penetration depth of HEA superconductors?" Are any other properties that reflect the basic physics of the superconductors of interest to determine? Do their superconducting properties look more like those of amorphous alloys or those of disordered crystalline alloys?

(6) Sophisticated materials-physics-intensive characterization of HEA superconductors remains sparse at the current time, at least in part due to the unavailability of single crystals. To our knowledge there are no current publications describing their fabrication, but as far as we can see there should not be a reason why their growth will be impossible. Can this be done? Angle resolved photoemission spectroscopy experiments would be of special interest for future characterization, for example, as would neutron scattering, in particular to investigate various of the phononic characteristics of HEA superconductors. Of great interest, for the type of material already available, would be the types of property characterization studies that have been performed on amorphous superconductors.

(7) What do the microstructures of HEA superconductors look like? How does phase separation happen at the nanoscale and microscale when stability limits are exceeded? What do grain boundaries, dislocations and stacking faults look like in HEA superconductors?

(8) Current studies have focused on creating single-phase HEA superconductors. How would the properties of HEA superconductors be impacted through precipitation of conventional or non-superconducting HEA phases within the superconducting matrix?

(9) What kinds of critical currents (Jc's) are attainable for HEA superconductor wires? Indeed can superconducting wires of meter-scale length be fabricated for HEAs?

(10) Although the mechanical and upper-critical field properties of HEA superconductors so far, based on limited information available, appear to be favorable for potential practical applications (although the Jc's remain uncharacterized) , will more focused, expert study find them to be worth considering as replacements for conventional superconducting alloys in real engineering systems or will they be limited to scientific interest only?

(11) Does the proven robustness of the superconductivity HEA materials under extreme pressure conditions promise unique applications for these materials?

(12) Can thin films of HEA superconductors be grown? How do their properties compare to those of the bulk materials?

(13) Bulk HEA superconductors are currently fabricated primarily through conventional arc-melting and fast although not ultrafast cooling. How might the compositional stability ranges or properties of HEA superconductors be impacted in alloys fabricated through different methods?

(14) Thus far, HEA superconductors appear to be far from potential magnetic instabilities. Are there any HEA superconductor systems where a magnetism-superconductivity connection, so significant in the copper oxide and iron pnictide superconductors, can be forged?

(15) Why have HEA superconductors with FCC structure not yet been found? Does this reflect the real exclusion of these structure types for HEA superconductors or only the statistics of small numbers?

(16) Studies that propose the existence of new superconducting HEAs and MEAs continue to be performed [51, 52]. What opportunities for the discovery of new superconducting materials do HEA superconductors provide? Can this family of materials provide a superconductor with a $T_C$ in the over 10 K range with excellent mechanical properties?

**Acknowledgements**

We thank Shu Cai and Jing Guo for their assistance in figure preparation. The work in China was supported by the National Key Research and Development Program of China (Grant No. 2017YFA0302900, 2016YFA0300300, and 2017YFA0303103), the NSF of China (Grant Numbers 11427805, U1532267, and 11604376), and the Strategic Priority Research

Program of the Chinese Academy of Sciences (Grant No. XDB25000000). The work at Princeton was supported by the Gordon and Betty Moore Foundation EPiQS initiative, Grant GBMF-4412.

Correspondence and requests for materials should be addressed to L. Sun (llsun@iphy.ac.cn) or R.J. Cava (rcava@Princeton.EDU)

Table 1 Overview with composition, superconducting transition temperature $T_C$, upper critical magnetic field $H_{C2}(0)$, structure and type of the high entropy alloy (HEA) superconductors. Type-A HEA superconductors crystallize on a BCC lattice, type-B HEA superconductors crystallize on an α-Mn lattice, type-C HEA superconductors crystallize on a CsCl lattice and type D crystallize on a HCP lattice.

| HEA-superconductor | $T_C$ (K) | $H_{C2}$ (0T) | Structure | Type | Reference |
|---|---|---|---|---|---|
| $(TaNb)_{0.7}(ZrHfTi)_{0.3}$ | 8.0 | 6.7 | BCC | A | [5] |
| $(TaNb)_{0.7}(ZrHfTi)_{0.33}$ | 7.8 | 7.8 | BCC | A | [5] |
| $Ta_{34}Nb_{33}Hf_{8}Zr_{14}Ti_{11}$ | 7.3 | 8.2 | BCC | A | [13] |
| $(TaNb)_{0.7}(ZrHfTi)_{0.4}$ | 7.6 | 8.4 | BCC | A | [5] |
| $(TaNb)_{0.7}(ZrHfTi)_{0.5}$ | 6.5 | 11.7 | BCC | A | [5] |
| $(TaNb)_{0.7}(ZrHfTi)_{0.84}$ | 4.5 | 9.0 | BCC | A | [5] |
| $(TaNb)_{67}(Hf)_{33}$ | 7.3 | - | BCC | A | [5] |
| $(TaNb)_{67}(HfZr)_{33}$ | 6.6 | - | BCC | A | [5] |
| $Nb_{67}(HfZrTi)_{33}$ | 9.2 | - | BCC | A | [5] |
| $(NbV)_{67}(HfZrTi)_{33}$ | 7.2 | - | BCC | A | [5] |
| $(TaV)_{67}(HfZrTi)_{33}$ | 4.0 | - | BCC | A | [5] |
| $(TaNb)_{67}(HfZrTi)_{33}$ | 7.3 | - | BCC | A | [5] |
| $(TaNbV)_{67}(HfZrTi)_{33}$ | 4.3 | - | BCC | A | [5] |
| $Nb_{25.2}Ta_{30.7}Ti_{21.3}Zr_{22.8}$ | 8.3 | 1.4 | BCC | A | [22] |
| $Nb_{22.1}Ta_{26.3}Ti_{16.6}Zr_{15.5}Hf_{19.5}$ | 7.1 | 2.0 | BCC | A | [22] |
| $Nb_{21.5}Ta_{18.1}Ti_{15.9}Zr_{14.4}Hf_{16.6}V_{13.5}$ | 5.1 | 2.0 | BCC | A | [22] |
| NbTaTiZrFe | 6.9 | - | BCC | A | [22] |
| NbTaTiZrGe | 8.4 | 1.3 | BCC | A | [22] |
| NbTaTiZrSiV | 4.3 | - | BCC | A | [22] |
| NbTaTiZrSiGe | 7.4 | 0.9 | BCC | A | [22] |
| $(ZrNb)_{0.2}(MoReRu)_{0.8}$ | 4.2 | - | BCC | B | [22] |
| $(ZrNb)_{0.1}(MoReRu)_{0.9}$ | 5.3 | 7.9 | BCC | B | [22] |
| $(HfTaWIr)_{0.6}Re_{0.4}$ | 1.9 | - | BCC+HCP | - | [21] |
| $(HfTaWIr)_{0.5}Re_{0.5}$ | 2.7 | - | BCC+HCP | - | [21] |
| $(HfTaWIr)_{0.4}Re_{0.6}$ | 4.0 | 4.7 | BCC | B | [21] |
| $(HfTaWIr)_{0.3}Re_{0.7}$ | 4.5 | - | BCC | B | [21] |
| $(HfTaWIr)_{0.2}Re_{0.8}$ | 5.7 | - | BCC | B | [21] |
| $(HfTaWPt)_{0.5}Re_{0.5}$ | 2.2 | - | BCC+HCP | - | [21] |
| $(HfTaWPt)_{0.4}Re_{0.6}$ | 4.4 | 5.9 | BCC | B | [21] |
| $(HfTaWPt)_{0.3}Re_{0.7}$ | 5.7 | - | BCC | B | [21] |
| $(HfTaWPt)_{0.2}Re_{0.75}$ | 6.1 | - | BCC | B | [21] |
| $(ScZrNbTa)_{0.65}(RhPd)_{0.35}$ | 9.3 | 10.7 | CsCl-type | C | [14] |
| $(ScZrNb)_{0.63}(RhPd)_{0.37}$ | 7.5 | 9.6 | CsCl-type | C | [14] |
| $(ScZrNb)_{0.62}(RhPd)_{0.38}$ | 6.4 | 8.9 | CsCl-type | C | [14] |
| $(ScZrNb)_{0.60}(RhPd)_{0.40}$ | 3.9 | 2.1 | CsCl-type | C | [14] |
| $Re_{0.56}Nb_{0.11}Ti_{0.11}Zr_{0.11}Hf_{0.11}$ | 4.4 | - | HCP | D | [29] |

Table 2 The superconducting temperatures $T_C$ for the isoelectronic substitutions with {Sc-Cr}, {Y-Mo}, and {Sc-Mo} mixtures in the pristine HEA superconductor $(TaNb)_{0.67}(HfZrTi)_{0.33}$ with $T_C \approx$ 7.6 K. The data were taken from Ref. [17].

| Substitution \ Original element | Ta | Nb | Hf | Zr | Ti |
|---|---|---|---|---|---|
| $Sc_{0.33}Cr_{0.67}$ | 5.6 | 4.4 | — | — | — |
| $Sc_{0.67}Cr_{0.33}$ | — | — | 7.5 | 7.4 | 7.6 |
| $Y_{0.33}Mo_{0.67}$ | 4.7 | 3.5 | — | — | — |
| $Y_{0.67}Mo_{0.33}$ | — | — | 7.6 | 6.7 | 7.5 |
| $Sc_{0.33}Mo_{0.67}$ | 4.4 | 2.9 | — | — | — |
| $Sc_{0.67}Mo_{0.33}$ | — | — | 7.5 | 6.6 | 7.5 |

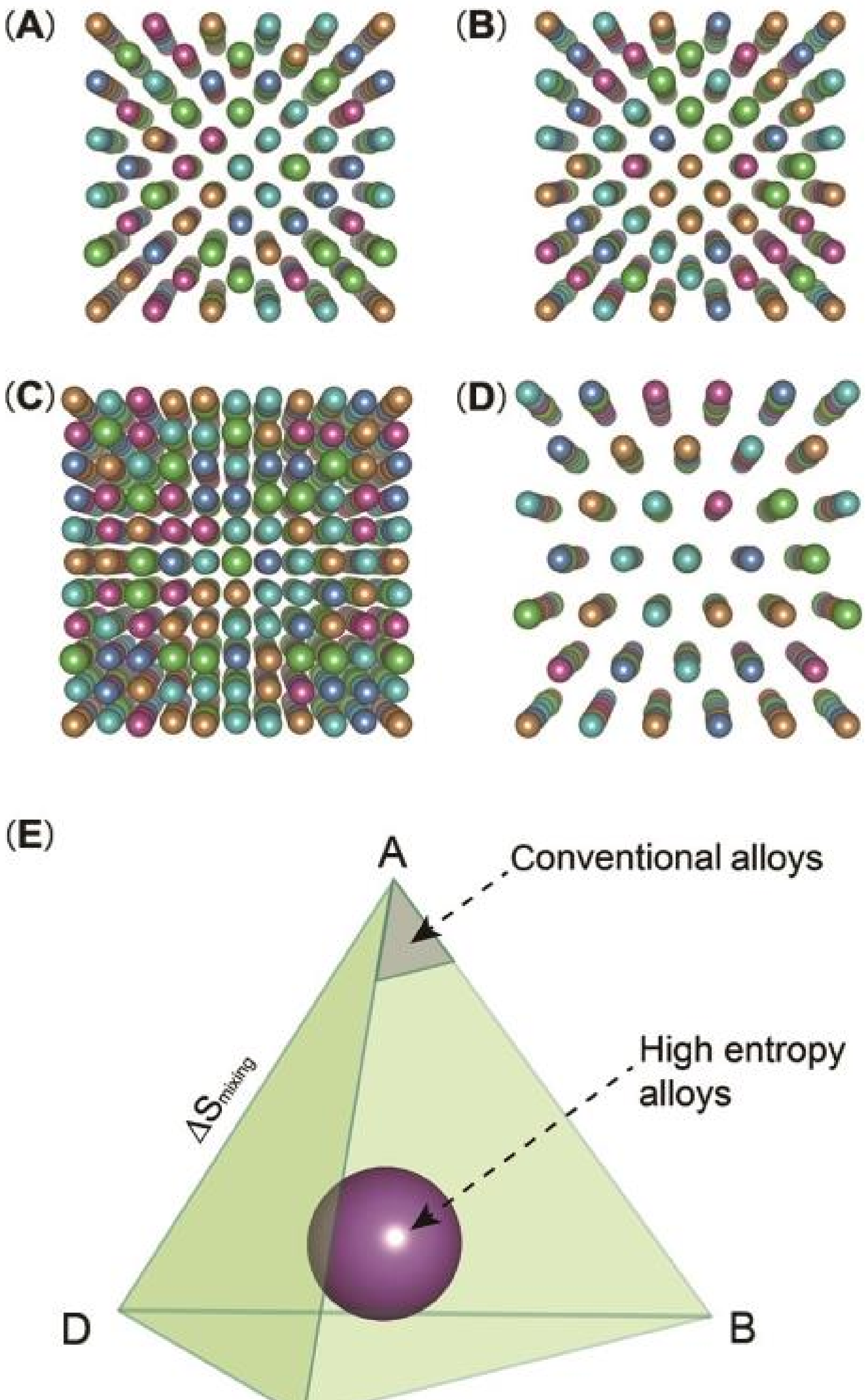

Figure 1 (A)-(D) Schematic representation of the BCC, CsCl, FCC and HCP lattices with randomly distributed atoms. (E) Schematic phase diagram of a multicomponent alloy system showing conventional and HEA phase regions.

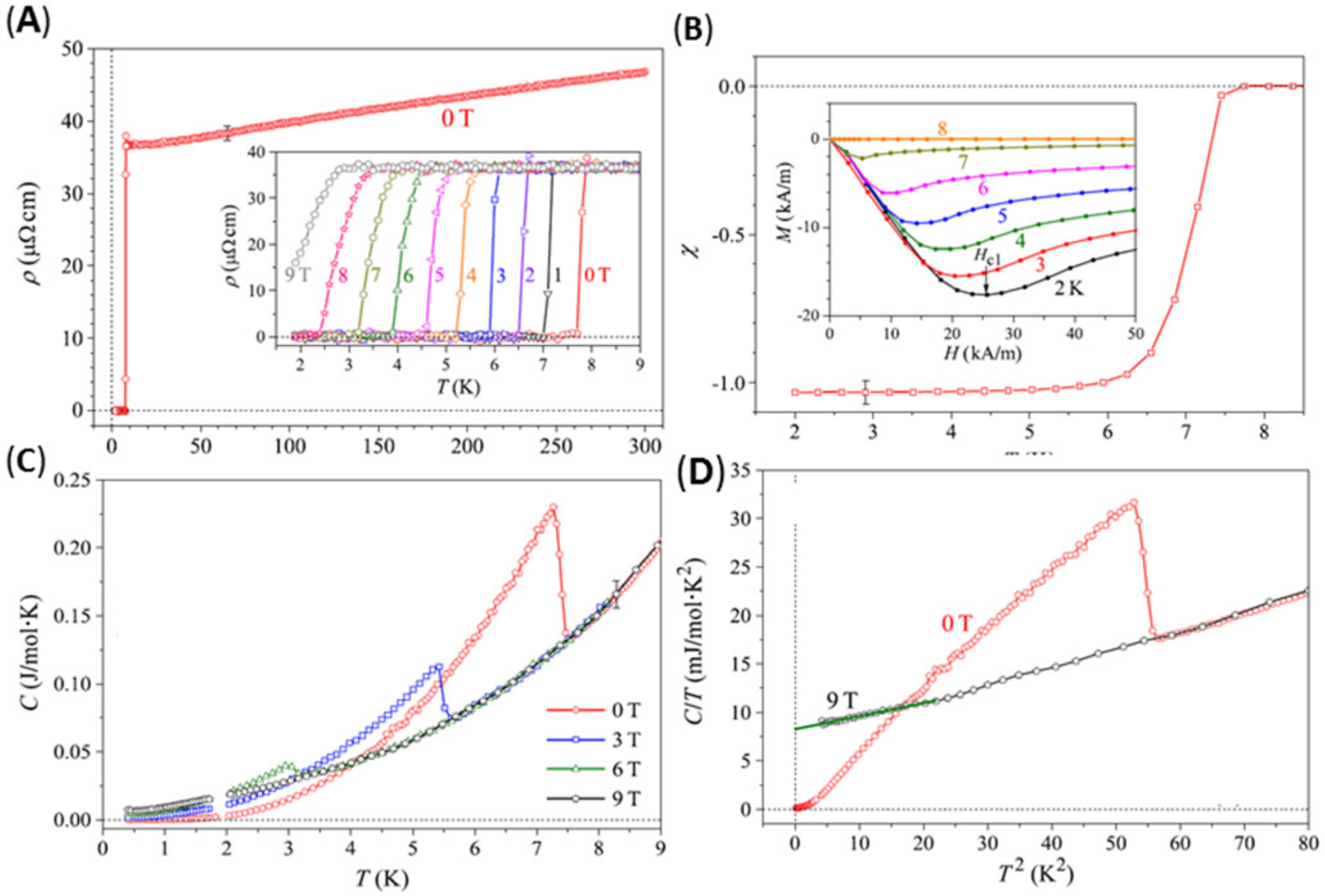

Figure 2 Ambient-pressure properties of the superconducting HEA $Ta_{34}Nb_{33}Hf_8Zr_{14}Ti_{11}$. (A) Resistivity as a function of temperature in zero magnetic field, showing the superconducting transition at 7.3 K. The inset displays the magnetic-field dependence of the resistivity in the region of the superconducting transition for fields up to 9 T. (B) The magnetic susceptibility in the region of the superconducting transition. The inset shows the isothermal magnetization in the low-field range. (C) Low-temperature specific heat C(T) for selected magnetic fields. (D) Specific heat in zero field and 9 T in a C/T versus $T^2$ plot. Reproduced with permission from [15].

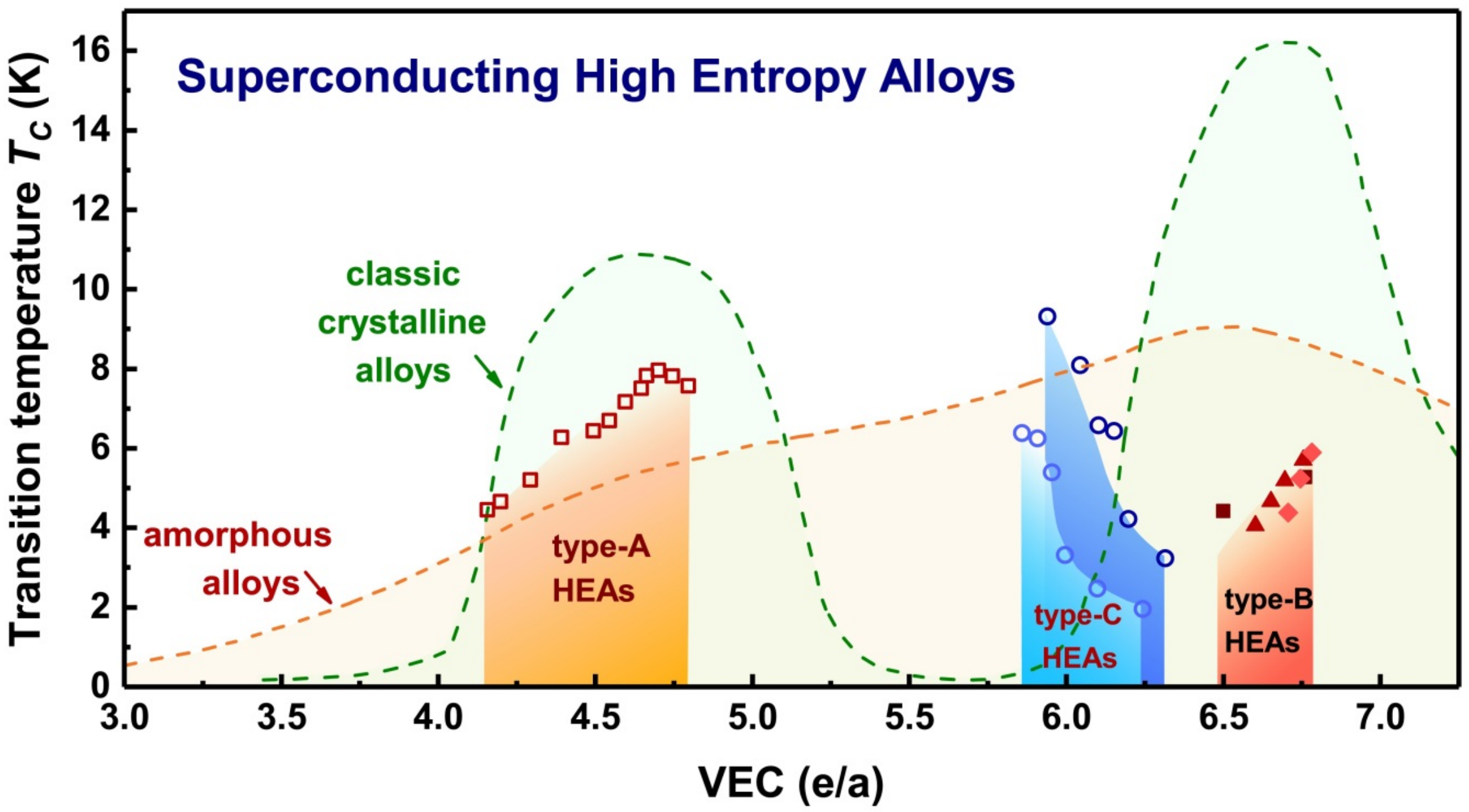

Figure 3 Valence electron count (VEC) dependence of the superconducting transition temperatures for type-A (orange region), type-B (cyan and blue regions) and type-C (red region) HEA superconductors compared to amorphous alloys (orange dashed line) and classic crystalline alloys (green dashed line). The data were taken from Ref.[7, 17, 23].

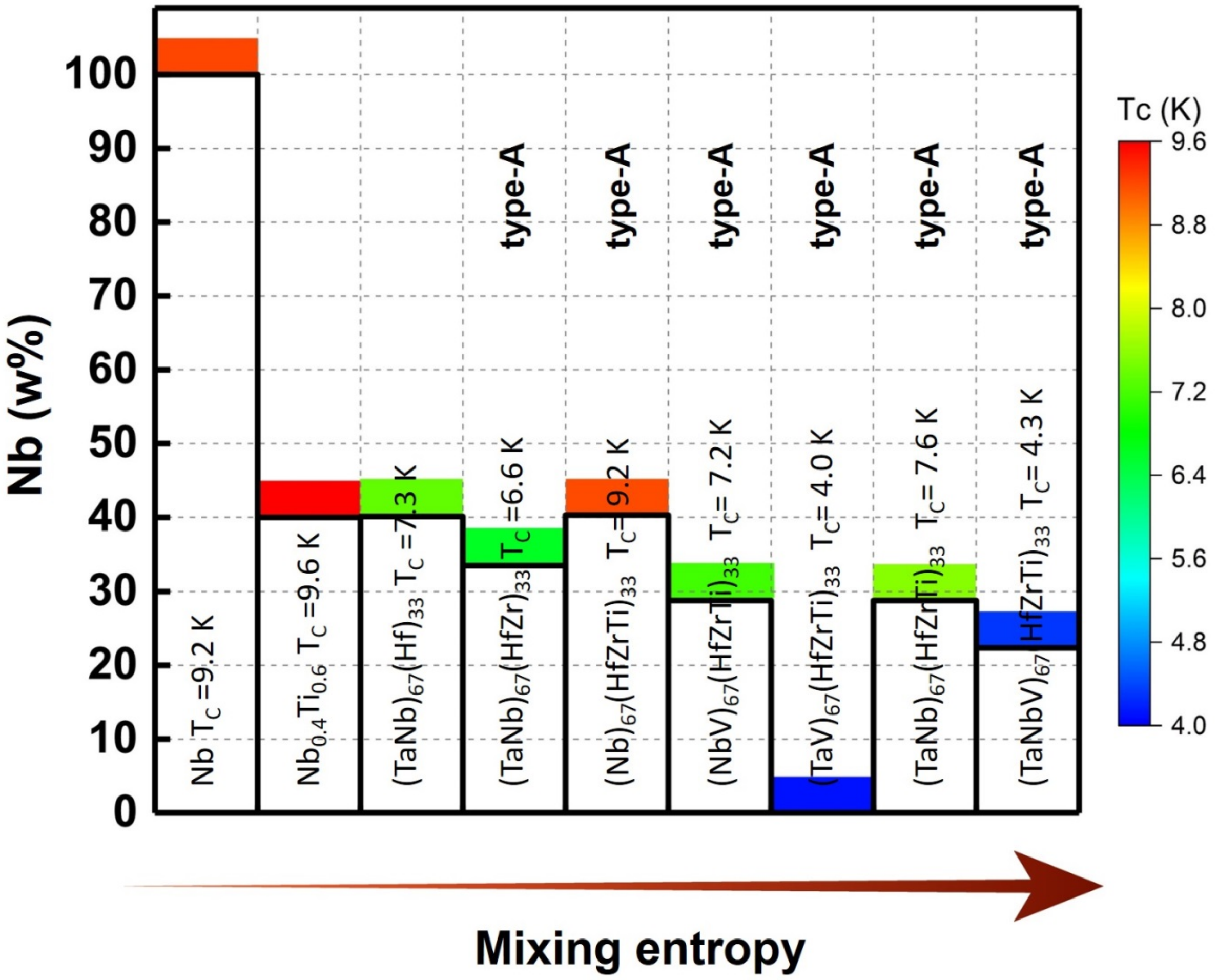

Figure 4 Plot of mixing entropy and Nb content versus superconducting transition temperature for HEA superconductors. The data were taken from Ref. [7].

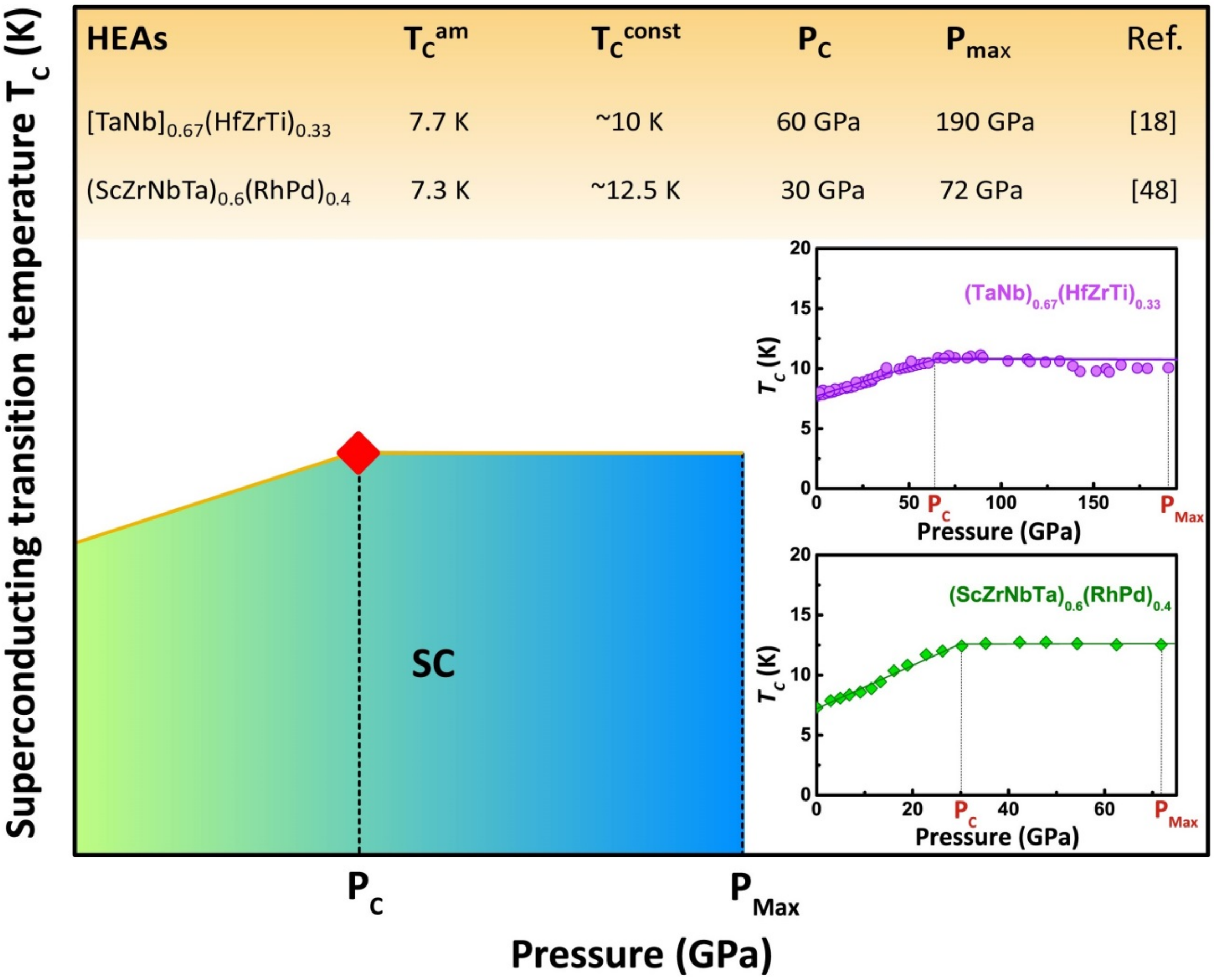

Figure 5 Schematic universal pressure-$T_C$ phase diagram of $(TaNb)_{0.67}(HfZrTi)_{0.33}$ (type-A HEA superconductor) and $(ScZrNbTa)_{0.6}(RhPd)_{0.4}$ (type-C HEA superconductor). $T_C^{am}$ and $T_C^{const}$ represent ambient-pressure $T_C$ and the $T_C$ which stays almost constant, respectively. $P_C$ and $P_{max}$ stand for the critical pressure where $T_C$ starts to be saturated and the highest pressure investigated, respectively. SC represents the superconducting phase. The insets display the real data.